\documentclass[prb,amsmath,amssymb,superscriptaddress,twocolumn]{revtex4}
\usepackage{graphicx}
\usepackage{bbm,color,ulem}
\usepackage{braket}
\usepackage{bbold}
\DeclareMathAlphabet{\mathpzc}{OT1}{pzc}{m}{it} 
\begin{document}
\title{Fidelity of a sequence of SWAP operations on a spin chain}
\author{Robert E.\ Throckmorton and S.\ Das Sarma}
\affiliation{Condensed Matter Theory Center and Joint Quantum Institute, Department of Physics, University of Maryland, College Park, Maryland 20742-4111 USA}
\date{\today}
\begin{abstract}
We consider the ``transport'' of the state of a spin across a Heisenberg-coupled spin chain via the use of
repeated SWAP gates, starting with one of two states---one in which the leftmost spin is down and the others
up, and one in which the leftmost two spins are in a singlet state (i.e., they are entangled), and the others
are again all up.  More specifically, we ``transport'' the state of the leftmost spin in the first case and
the next-to-leftmost spin in the second to the other end of the chain, and then back again.  We accomplish our
SWAP operations here by increasing the exchange coupling between the two spins that we operate on from a base
value $J$ to a larger value $J_\text{SWAP}$ for a time $t=\pi\hbar/4J_\text{SWAP}$.  We determine the fidelity
of this sequence of operations in a number of situations---one in which only nearest-neighbor coupling exists
between spins and there is no magnetic dipole-dipole coupling or noise (the most ideal case), one in which we
introduce next-nearest-neighbor coupling, but none of the other effects, and one in which all of these effects
are present.  In the last case, the noise is assumed to be quasistatic, i.e., the exchange couplings are each
drawn from a Gaussian distribution, truncated to only nonnegative values.  We plot the fidelity as a function
of $J_\text{SWAP}$ to illustrate various effects, namely crosstalk due to coupling to other spins, as well as
noise, that are detrimental to our ability to perform a SWAP operation.  Our theory should be useful to the
ongoing experimental efforts in building semiconductor-based spin quantum computer architectures.
\end{abstract}
\maketitle
\section{Introduction}
The ability to transfer the state of one qubit to another is critical in quantum computation.  A key part
of this transfer of qubit states is the ability to perform a SWAP gate, even just between two neighboring qubits.
Aside from being useful on its own for this purpose, it is also useful for performing quantum teleportation, as
it would allow one to transfer entanglement from one qubit to another.  A number of experimental groups have
demonstrated the ability to perform such gates in electron spin-based quantum dot systems.  One experiment\cite{KandelNature2019}
has demonstrated the ability to perform a SWAP gate with a fidelity of around $90\%$, while another\cite{SigillitoNPJQI2019}
has implemented SWAP gates for spin eigenstates (i.e., it can only swap purely up or purely down spins) with a
fidelity of $98\%$ and for arbitrary states with a fidelity of $84\%$.  Both of these experiments were performed
on semiconductor-based electron spin qubits.  In addition, there has been experimental and theoretical interest
in two-qubit gates in general\cite{FowlerPRA2004,ShulmanScience2012,VeldhorstNature2015,ZajacScience2018,WatsonNature2018,HuangNature2019,XuePRX2019}
due to the fact that the ability to perform a two-qubit entangling gate is essential to universal quantum
computation.  The current theoretical work is on solid state spin qubits in semiconductor-based scalable
platforms, where the exchange coupling between neighboring localized electrons is typically used to carry
out SWAP operations.  The main physics we theoretically address here is the fidelity of SWAP-induced spin
transport through a sequence of qubits in the presence of non-ideal effects invariably present in real physical
systems.

An essential part of the development of a quantum computer is the ability to perform any gate with a fidelity
of at least $99.9\%$ (and much higher), which is the threshold above which error-correcting codes may be implemented.  We are
therefore interested in characterizing the fidelity of SWAP operations in a model system.  We will consider
here a chain of electronic spins with Heisenberg exchange coupling.  We will consider three cases---one in which
there is only nearest-neighbor exchange coupling and no noise, one in which we add next-nearest-neighbor exchange
(but no noise) as well, and finally one in which we also add in the magnetic dipole-dipole interaction (even though it would
be very weak in actual experimental systems) and noise in the exchange couplings.  For each case, we will consider
two initial conditions, one in which the leftmost spin is initialized in the down state and the rest in the up
state (i.e., $\ket{\psi_{0,1}}=\ket{\downarrow\uparrow\uparrow\cdots\uparrow}$), and one in which the leftmost two
spins are prepared in a singlet state $\ket{S}=\frac{1}{\sqrt{2}}(\ket{\uparrow\downarrow}-\ket{\downarrow\uparrow})$
and the rest are all prepared in the up state (i.e., $\ket{\psi_{0,2}}=\ket{S}\otimes\ket{\uparrow\cdots\uparrow}$).
We then consider a sequence of SWAP operations, each of which is implemented as follows.  We increase the nearest-neighbor
coupling between the two spins that we perform this operation on from a base value $J$ to a larger value $J_\text{SWAP}$ and maintain
said value for a time $t=\pi\hbar/4J_\text{SWAP}$.  We assume for the purposes of this work that the exchange coupling can
be changed instantaneously, so that the Hamiltonian may be treated as piecewise constant in time.  In actual experiments,
however, there is always a finite ramp up or down time, but we do not expect this to affect our results significantly
in light of similar work on a single qubit that considers the effects of such a ramp up or down\cite{ThrockmortonPRB2019}.
In the most ideal case, i.e., there is no interaction of the two spins involved with other spins and no noise in the
exchange coupling, this performs a perfect SWAP operation.  We will assume that the next-nearest-neighbor exchange
coupling of the two spins involved to other spins, if it is included, increases in proportion to the nearest-neighbor
coupling.  The sequence of operations is as follows.  In the case in which we start in the state, $\ket{\psi_{0,1}}$,
we ``move'' the down spin all the way to the right, and then back again.  The sequence for the case in which we start
in the state, $\ket{\psi_{0,2}}$ is similar---we ``move'' the state of the right spin in the entangled pair all the way
to the right, and then back again.  In this latter case, the SWAP operation has the effect of transferring entanglement
with the leftmost spin to spins further right on the chain.  We then numerically determine the fidelity of this sequence
of operations, defined as the probability that the system, when measured, will be in the state that it would be in if
the intended sequence was performed without errors, as a function of $J_\text{SWAP}$.  In the case that we have noise
in the system, we assume that the noise is quasistatic, with all exchange couplings chosen from a Gaussian distribution with mean
$J_0$ and standard deviation $\sigma_J$, which we will call the strength of the noise, truncated so that all exchange
couplings are nonnegative.  In all cases, we will consider chains of $4$ and $6$ spins.  Larger number of qubits ($>6$)
can be easily studied using our technique, but the current experiments are restricted only to a few spins, and therefore,
we restrict our work to $6$ spins at most.  Larger number of spins would only suppress the calculated fidelity under
the same conditions.

We find that, even in the case with only nearest-neighbor coupling, there is a loss of fidelity, due to crosstalk
from the neighboring spins, but the fidelity approaches unity as $J_\text{SWAP}$ increases due to the shorter time
that this larger exchange coupling must be maintained, which gives the crosstalk-inducing terms too little time to
have a significant effect on the SWAP operation.  If next-nearest-neighbor coupling is included, however, then the
infidelity ($1-F$, where $F$ is the fidelity) saturates to a nonzero value, even without noise.  As noted earlier,
we increase the next-nearest-neighbor exchange couplings for the two spins involved in the SWAP operation in proportion
to the nearest-neighbor coupling, so that there is always significant crosstalk in this case.  We then consider adding
the dipole-dipole interaction and quasistatic noise.  We find that, for large values of $J_\text{SWAP}$, the fidelity
decreases monotonically as the noise strength increases, though the relationship becomes inverted for some smaller
values of $J_\text{SWAP}$.  We also indicate about how much noise the system is allowed to have to achieve the values found
in the known experiments, as well as to reach the $99.9\%$ threshold.

The rest of our paper is organized as follows.  We introduce the Hamiltonian for our system in Sec.\ II and introduce
the problem that we consider in more detail.  We review how to implement a SWAP gate in the system under consideration
and give our results for each case that we consider in Sec.\ III.  We then conclude in Sec.\ IV.

\section{Hamiltonian}
We consider here a chain of spins coupled with both nearest- and next-nearest-neighbor Heisenberg exchange
couplings and a magnetic dipole-dipole interaction.  The Hamiltonian describing this system is
\begin{eqnarray}
H&=&-\sum_{j=1}^{L-1}J_j\vec{\sigma}_j\cdot\vec{\sigma}_{j+1}-\sum_{j=1}^{L-2}J'_j\vec{\sigma}_j\cdot\vec{\sigma}_{j+2} \cr
&-&\frac{\mu_0\hbar^2}{16\pi a^3}\sum_{ij}\frac{1}{|i-j|^3}(2\sigma_{i,z}\sigma_{j,z}-\sigma_{i,x}\sigma_{j,x}-\sigma_{i,y}\sigma_{j,y}), \nonumber \\
\end{eqnarray}
where $\vec{\sigma}_j$ is the vector of Pauli matrices acting on spin $j$, $L$ is the number of spins, $a$ is
the distance between two adjacent spins and $\mu_0$ is the magnetic permeability of free space.  We allow each
of the exchange couplings to depend on position to allow the addition of quasistatic noise (mathematically
identical to disorder).

We will consider two problems.  First, let us initialize the system so that the leftmost spin is down,
but all other spins are up (i.e., the initial state is $\ket{\psi_{0,1}}=\ket{\downarrow\uparrow\uparrow\cdots\uparrow}$).
We then apply a series of SWAP gates to ``move'' the down spin all the way to the right, and then all
the way back to the left.  We will then calculate the fidelity of this sequence of moves, defined as
the probability that the state that we measure the system in is that which we would obtain ideally
(in this case, this would just be the initial state):
\begin{equation}
F=\left |\bra{\psi_0}U^\dag R\ket{\psi_0}\right |^2,
\end{equation}
where $\ket{\psi_0}$ is the initial state, $R$ is the ideal sequence of operations (in our case, just the identity
operation), and $U$ is the actual sequence of operations performed under the influence of error-causing effects.
When we consider the effects of noise, we will average this fidelity over different realizations of noise and report
the noise-averaged fidelity $\bar{F}$.  The second problem that we will consider is similar---we initialize the
system so that the leftmost two spins are in a singlet state, $\ket{S}=\frac{1}{\sqrt{2}}(\ket{\uparrow\downarrow}-\ket{\downarrow\uparrow})$,
while the rest are up (i.e., the full initial state is now $\ket{\psi_{0,2}}=\frac{1}{\sqrt{2}}(\ket{\uparrow\downarrow}-\ket{\downarrow\uparrow})\otimes\ket{\uparrow\uparrow\cdots\uparrow}$),
and then consider a chain of SWAP operations that ``move'' the state of the second spin all the way to the right,
and then back again to the second spin.  The effect of each of these SWAP operations in this case is to transfer
the second spin's entanglement with the first to each of the other spins to its right.

\section{Results}
We now present the results that we obtain.  Before doing so, however, we first review how we perform SWAP gates
in our model in the most ideal case.  We consider a system consisting of just two exchange-coupled spins (we ignore
the dipole-dipole interaction),
\begin{equation}
H=-J\vec{\sigma}_1\cdot\vec{\sigma}_2.
\end{equation}
If we write this in matrix form in the basis, $(\ket{\uparrow\uparrow},\ket{\uparrow\downarrow},\ket{\downarrow\uparrow},\ket{\downarrow\downarrow})$,
we get
\begin{equation}
H=\begin{bmatrix}
J & 0 & 0 & 0 \\
0 & -J & 2J & 0 \\
0 & 2J & -J & 0 \\
0 & 0 & 0 & J
\end{bmatrix}.
\end{equation}
Note that this is a block diagonal matrix; this allows us to use the identity for $2\times 2$ Hermitian
matrices,
\begin{equation}
e^{i\vec{v}\cdot\vec{\sigma}}=\cos{v}\mathbb{1}+i\hat{v}\cdot\vec{\sigma}\sin{v},
\end{equation}
where $v=|\vec{v}|$ and $\hat{v}=\vec{v}/v$, to obtain the corresponding time evolution operator,
\begin{eqnarray}
U(t)&=&e^{-iHt/\hbar} \cr
&=&\begin{bmatrix}
e^{-i\tau} & 0 & 0 & 0 \\
0 & e^{i\tau}\cos\left (2\tau\right ) & -ie^{i\tau}\sin\left (2\tau\right ) & 0 \\
0 & -ie^{i\tau}\sin\left (2\tau\right ) & e^{i\tau}\cos\left (2\tau\right ) & 0 \\
0 & 0 & 0 & e^{-i\tau}
\end{bmatrix}, \nonumber \\
\end{eqnarray}
where $\tau=\frac{J}{\hbar}t$.  If we now let $t=\pi\hbar/4J$, this becomes
\begin{equation}
U\left (\frac{\pi\hbar}{4J}\right )=e^{-i\pi/4}\begin{bmatrix}
1 & 0 & 0 & 0 \\
0 & 0 & 1 & 0 \\
0 & 1 & 0 & 0 \\
0 & 0 & 0 & 1
\end{bmatrix}=e^{-i\pi/4}\text{SWAP}.
\end{equation}
We see that this Hamiltonian implements a SWAP operation up to an (unimportant) overall phase factor.  Even
with more than two qubits, this phase would only appear as an unmeasurable global phase on the wave function,
and thus will have no effect on the fidelity of the operation on its own.  The fidelity is thus reduced only
by crosstalk with the other qubits and by noise in the exchange coupling.

\subsection{Noiseless case}
Let us first consider the noiseless case.  In this case, all $J_j=J$ and $J'_j=J'$.  We will consider in this
subsection cases in which $J'=0$ (i.e., nearest-neighbor coupling only) and $J'=0.01J$.  We also assume that
the ``base'' value of $J$ is $150\text{ kHz}$, or $\frac{1}{1,000}$ the value used in the simulations in Ref.\
\onlinecite{KandelNature2019}; therefore, the largest value of $J_\text{SWAP}$ that we consider here is exactly
the value used in this reference.  This results in the dipole-dipole interaction being negligible (on the order
of tens of Hz for nearest neighbors for a typical experimental system), so we will neglect it for now.  When we
perform a SWAP operation, we assume that the next-nearest-neighbor exchange couplings involving the two spins
undergoing the SWAP increase in proportion to the nearest-neighbor coupling.  As an example, if we perform a
SWAP operation on spins $2$ and $3$, then the couplings between spins $1$ and $3$ and between $2$ and $4$ will
also increase proportionally.  This assumed proportionality between the nearest-neighbor ($J$) and the next-nearest
neighbor ($J'$) spin-spin exchange coupling is reasonable for fixed localized spins with the proportionality
constant typically being small, depending on the inter-qubit spacings in the system.  If the exchange coupling
falls off exponentially with interqubit spacing, which is approximately the situation for quantum dot or donor
based spin qubit architectures, then $J'=0.01J$, as assumed in our calculations, is most likely a very optimistic
estimate for the next-nearest-neighbor spin coupling strength hindering the SWAP operation.  If we assume exponential
localization of the electronic wave function within a quantum dot, so that $J'/J\sim e^{-d/a}$, where $d$ is the
distance between quantum dots and $a$ is the width of the dots, then we would find that $J'\approx 0.13J$ in the
experiment of Ref.\ \onlinecite{KandelNature2019}, resulting in a much larger next-nearest-neighbor coupling.
A visual estimate of the relevant dimensions yields $d\approx 200\text{ nm}$ and $a\approx 100\text{ nm}$.
On the other hand, if we assume that the wave function is Gaussian, so that $J'/J\sim e^{-3(d/a)^2}$, then we
find that the next-nearest-neighbor exchange coupling is much smaller, $J'\approx 6.14\times 10^{-6}J$.  Therefore,
we see that the magnitude of $J'$ relative to $J$ varies depending on the details of how the wave function
falls off with distance.  Our assumption of $J'=0.01J$ here falls between these two values, so we expect that
it is a fair estimate of the actual value.

\begin{figure*}[htb]
	\centering
		\includegraphics[width=\columnwidth]{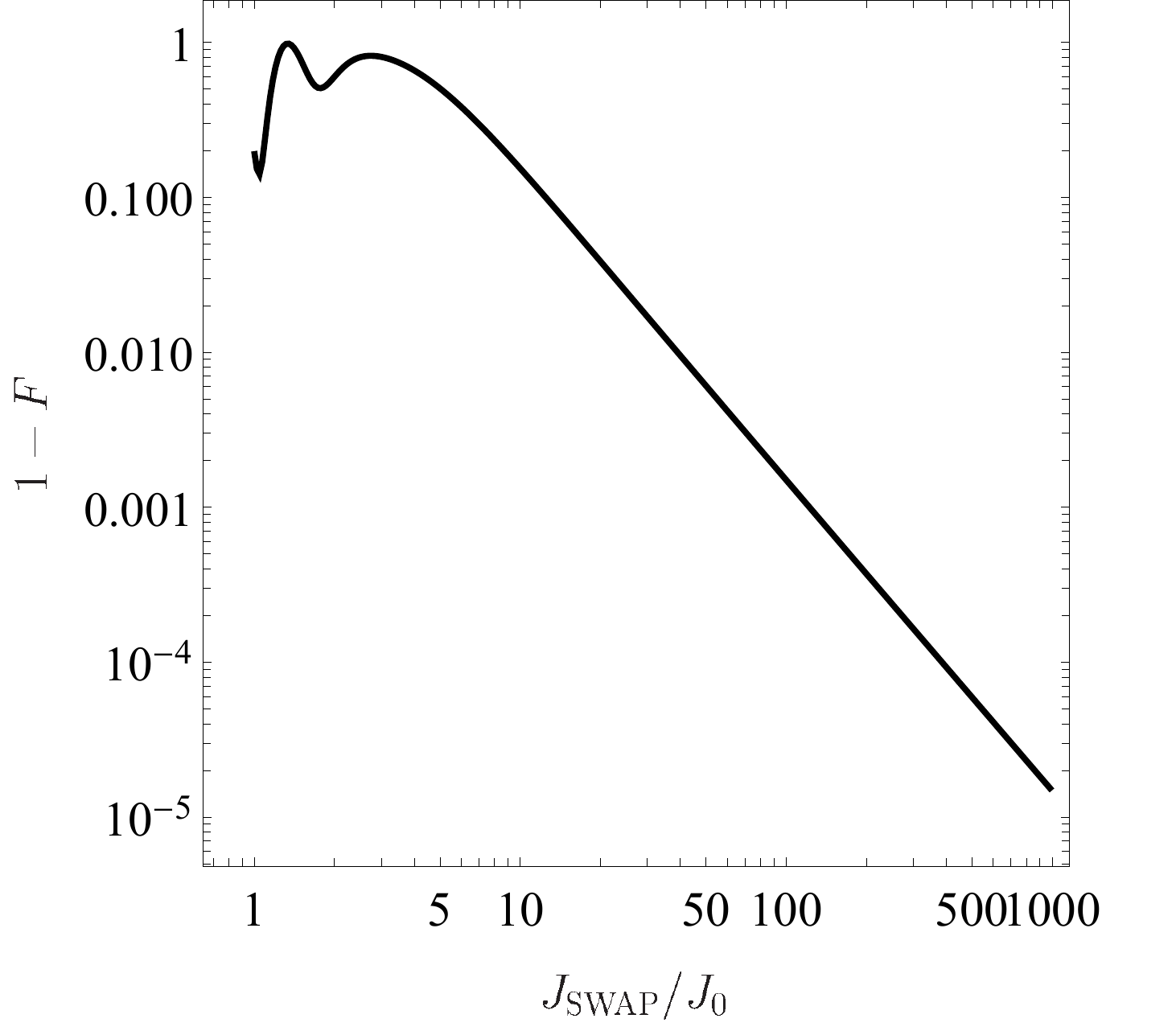}
		\includegraphics[width=\columnwidth]{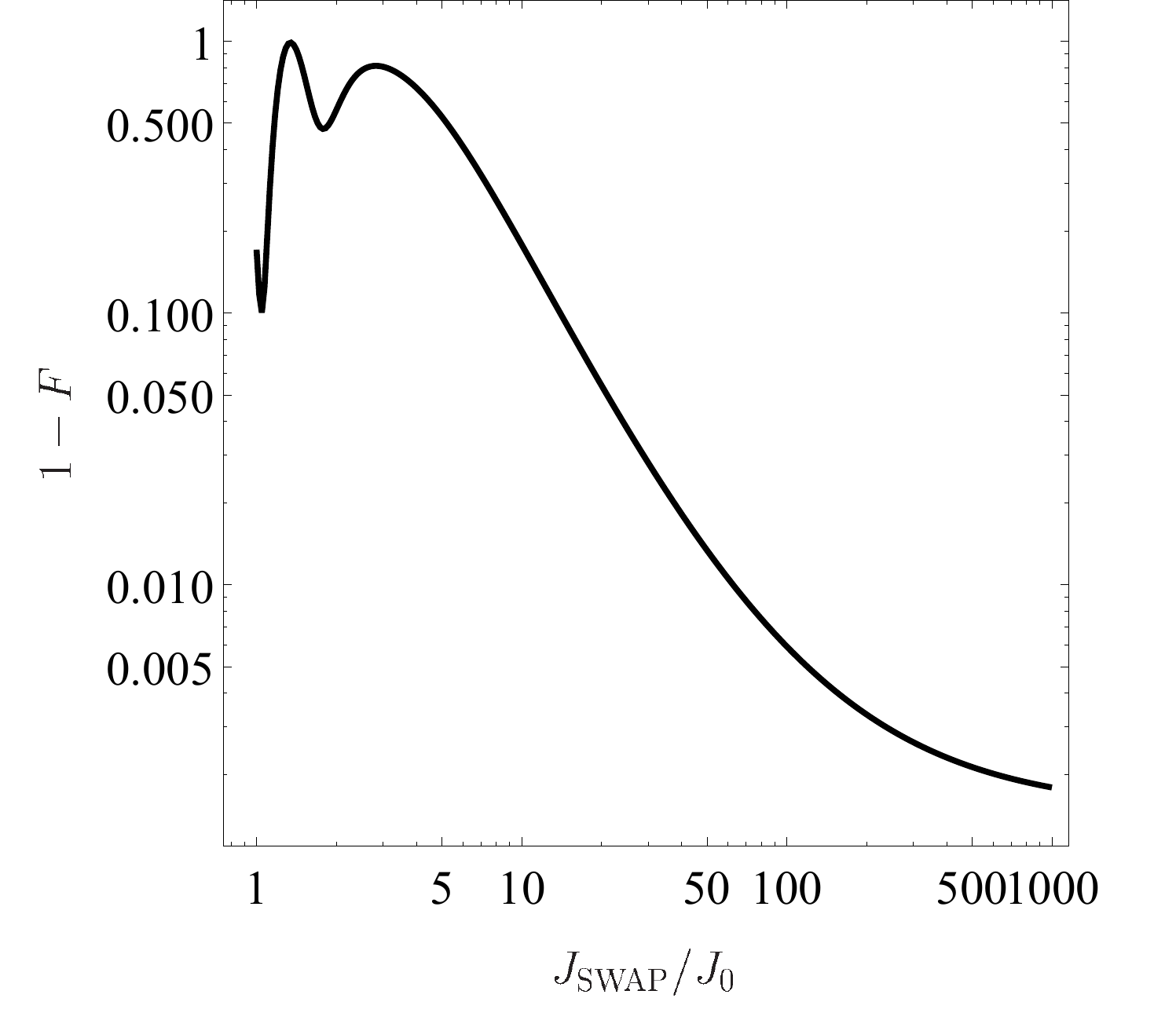}
	\caption{Plot of infidelity $1-F$ of ``transporting'' a down spin from left to right and back again as a
	function of $J_\text{SWAP}/J$ for $4$ spins in the absence of noise and dipole-dipole coupling.  The system
	is initialized in the state, $\ket{\psi_{0,1}}=\ket{\downarrow\uparrow\uparrow\uparrow}$.  The plot on the left assumes
	a nearest-neighbor coupling only, i.e., $J'=0$, while that on the right assumes a next-nearest-neighbor coupling
	$J'=0.01J$.}
	\label{fig:Nosj_Plot_4}
\end{figure*}

\begin{figure*}[htb]
	\centering
		\includegraphics[width=\columnwidth]{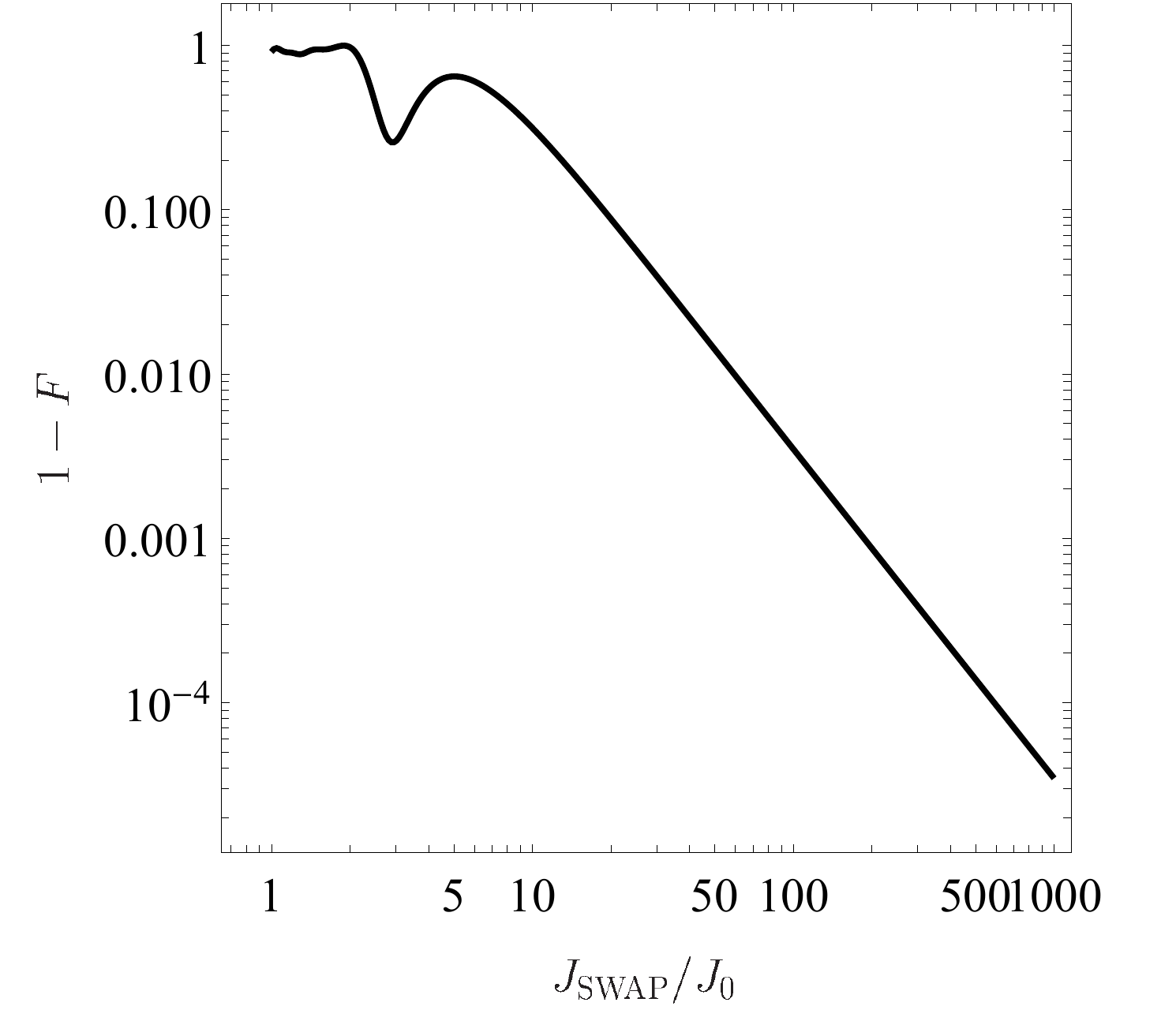}
		\includegraphics[width=\columnwidth]{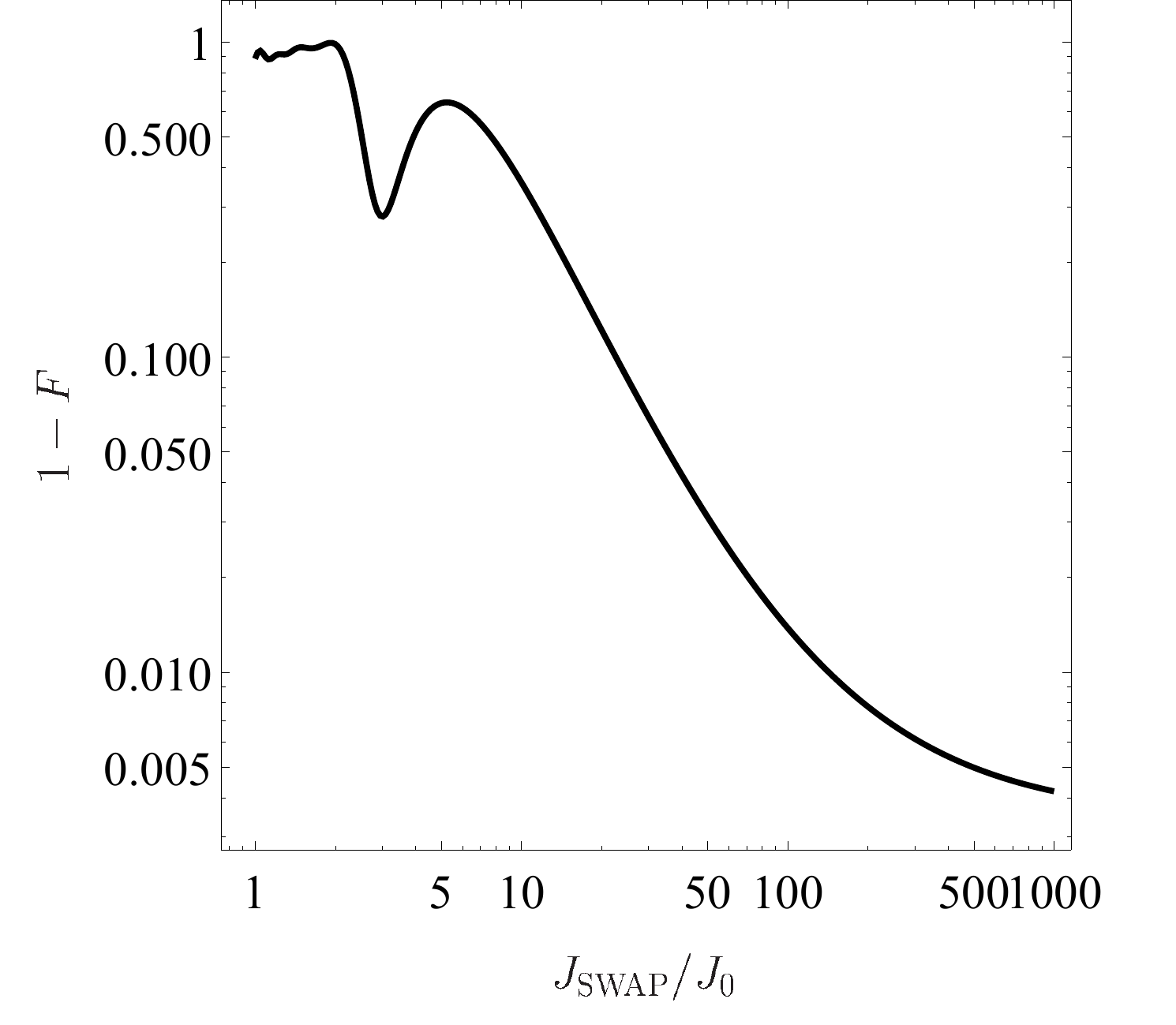}
	\caption{Plot of infidelity $1-F$ of ``transporting'' a down spin from left to right and back again as a
	function of $J_\text{SWAP}/J$ for $6$ spins in the absence of noise and dipole-dipole coupling.  The system
	is initialized in the state, $\ket{\psi_{0,1}}=\ket{\downarrow\uparrow\uparrow\uparrow\uparrow\uparrow}$.  The plot on the left assumes
	a nearest-neighbor coupling only, i.e., $J'=0$, while that on the right assumes a next-nearest-neighbor coupling
	$J'=0.01J$.}
	\label{fig:Nosj_Plot_6}
\end{figure*}

\begin{figure*}[htb]
	\centering
		\includegraphics[width=\columnwidth]{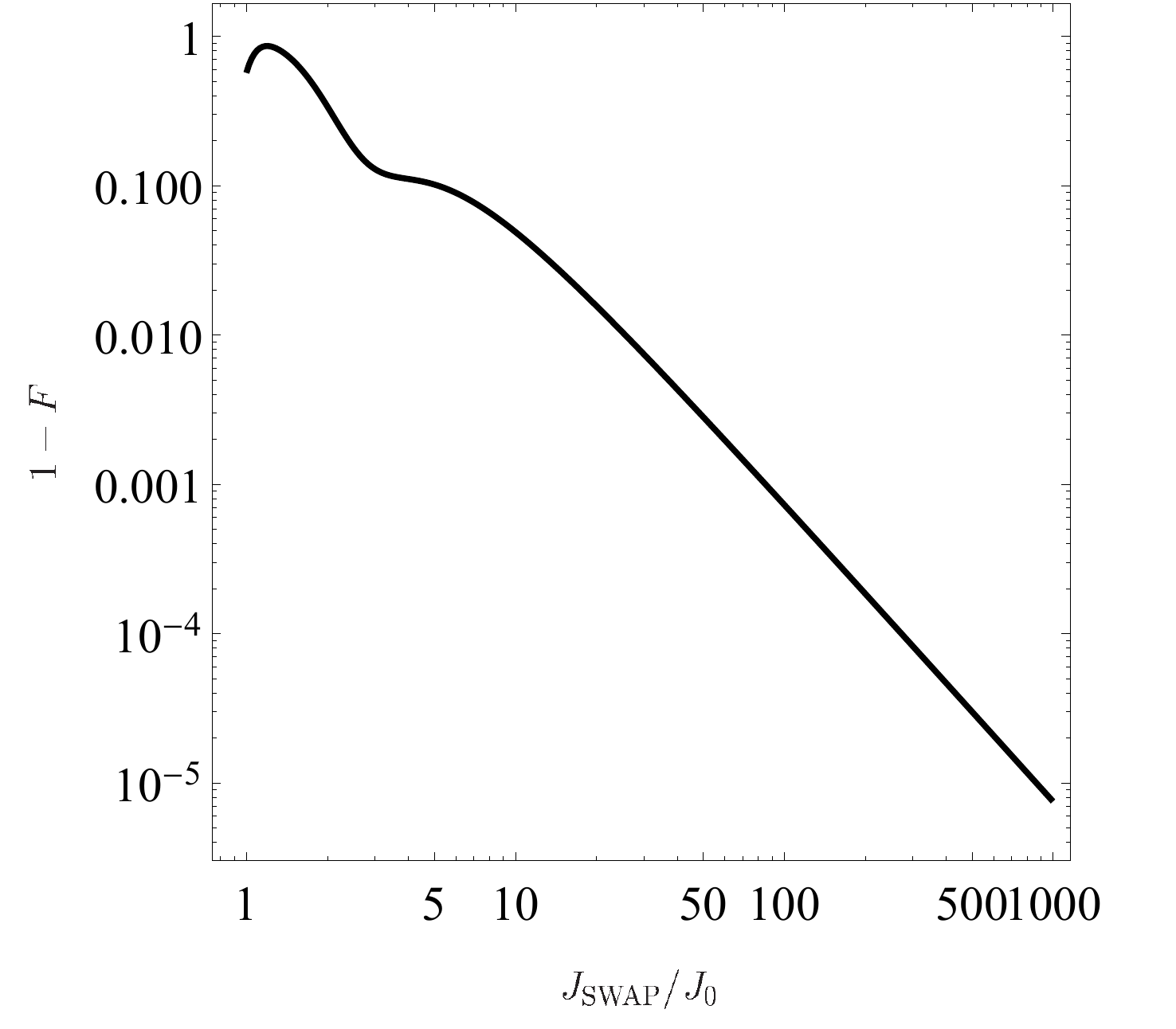}
		\includegraphics[width=\columnwidth]{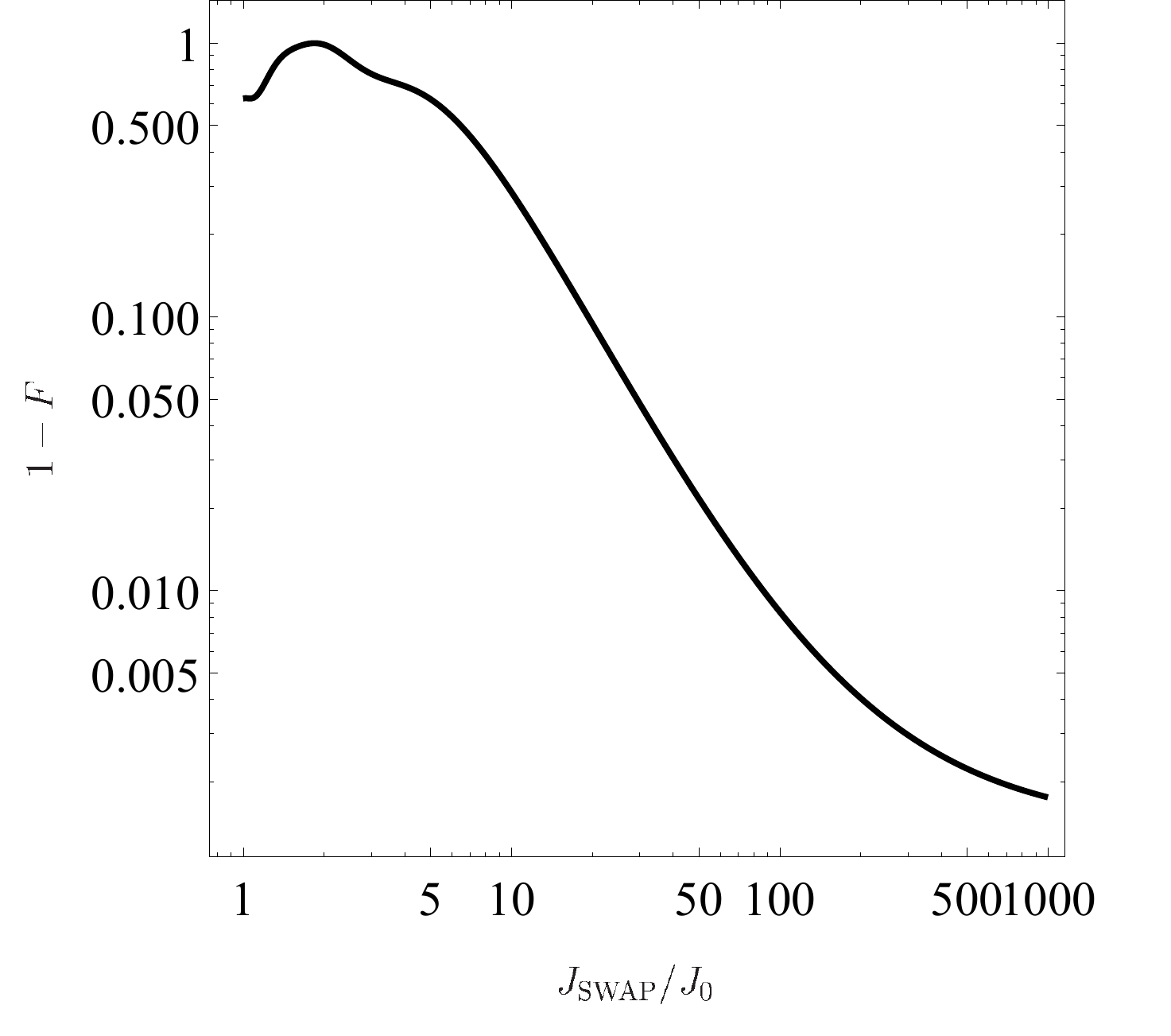}
	\caption{Plot of infidelity $1-F$ of ``transporting'' one of two entangled spins from left to right and back
	again as a function of $J_\text{SWAP}/J$ for $4$ spins in the absence of noise and dipole-dipole coupling.
	The system is initialized in the state, $\ket{\psi_{0,2}}=\ket{S}\otimes\ket{\uparrow\uparrow}$, where
	$\ket{S}=\frac{1}{\sqrt{2}}(\ket{\uparrow\downarrow}-\ket{\downarrow\uparrow})$.  The plot on the left assumes
	a nearest-neighbor coupling only, i.e., $J'=0$, while that on the right assumes a next-nearest-neighbor coupling
	$J'=0.01J$.}
	\label{fig:Nosj_Singlet_Plot_4}
\end{figure*}

\begin{figure*}[htb]
	\centering
		\includegraphics[width=\columnwidth]{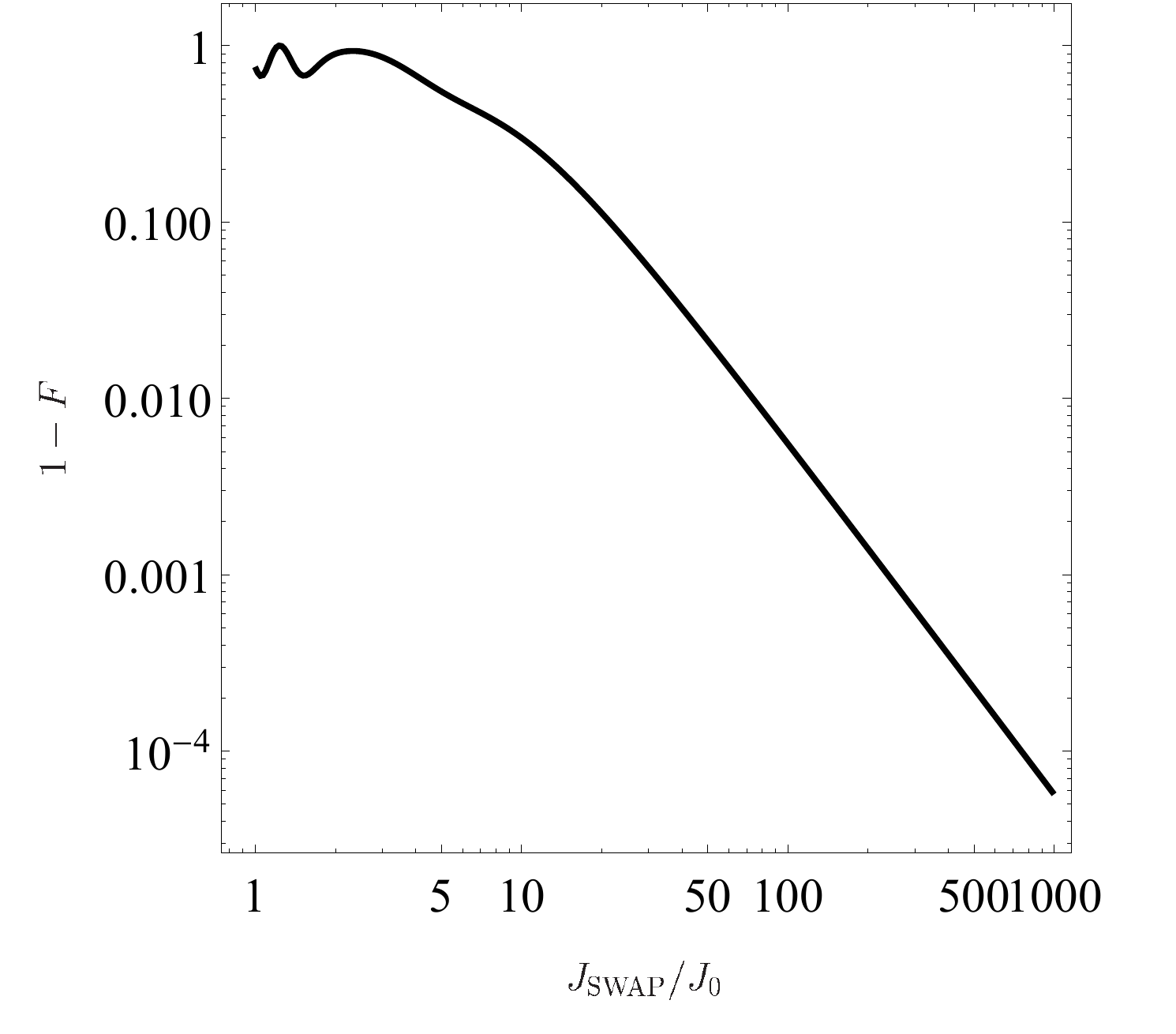}
		\includegraphics[width=\columnwidth]{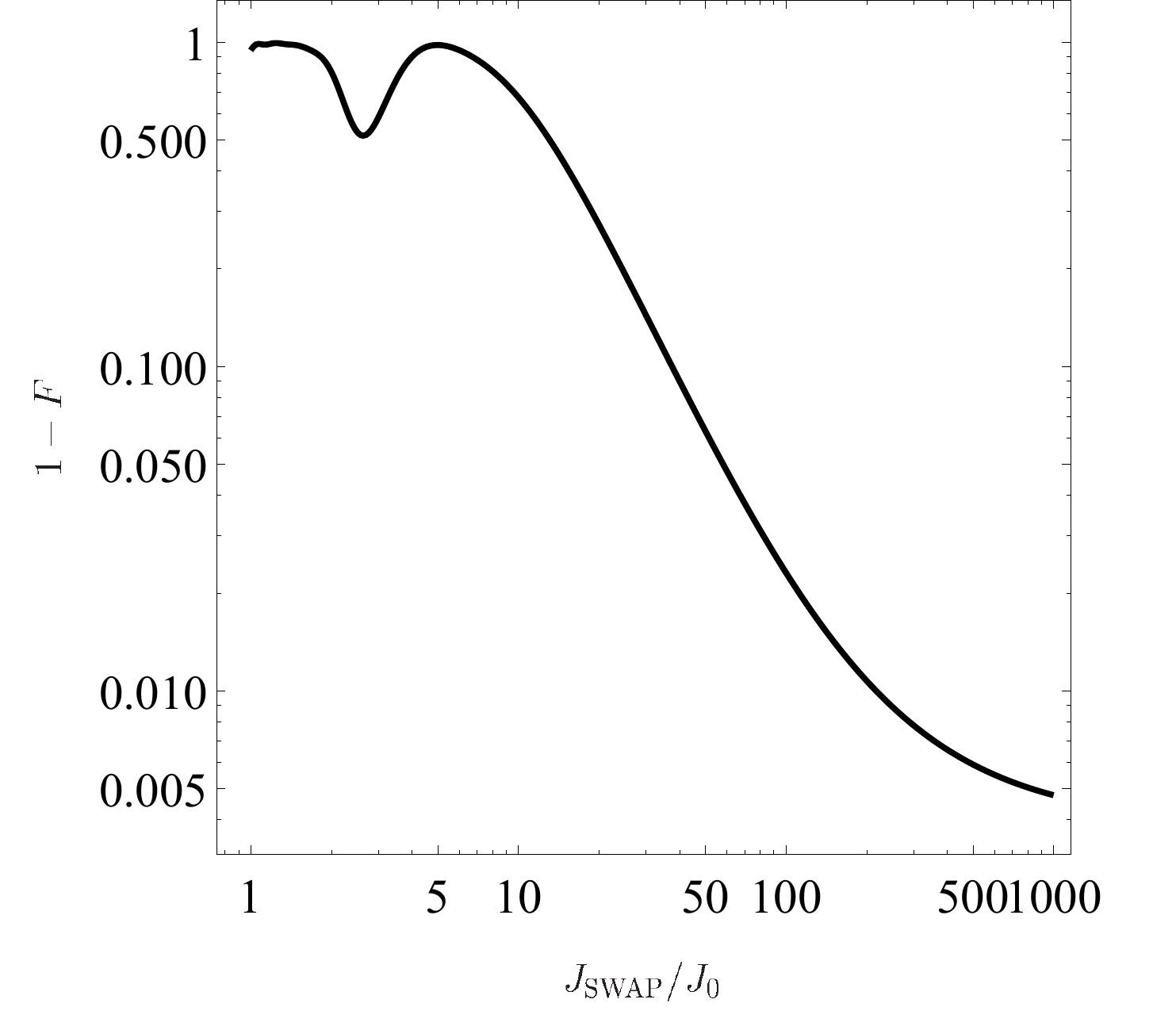}
	\caption{Plot of infidelity $1-F$ of ``transporting'' one of two entangled spins from left to right and back
	again as a function of $J_\text{SWAP}/J$ for $6$ spins in the absence of noise and dipole-dipole coupling.
	The system is initialized in the state, $\ket{\psi_{0,2}}=\ket{S}\otimes\ket{\uparrow\uparrow\uparrow\uparrow}$,
	where $\ket{S}=\frac{1}{\sqrt{2}}(\ket{\uparrow\downarrow}-\ket{\downarrow\uparrow})$.  The plot on the left
	assumes a nearest-neighbor coupling only, i.e., $J'=0$, while that on the right assumes a 	next-nearest-neighbor
	coupling $J'=0.01J$.}
	\label{fig:Nosj_Singlet_Plot_6}
\end{figure*}

We present our results for the case in which we start in the state, $\ket{\psi_{0,1}}$, in Figs.\ \ref{fig:Nosj_Plot_4}
and \ref{fig:Nosj_Plot_6} for values of $J_\text{SWAP}/J$ from $1$ to $1,000$ and for chains of $4$ and
$6$ spins.
We see that, even in the absence of noise, there is error in the overall SWAP operations, especially for smaller
values of $J_\text{SWAP}$.  This is due to crosstalk, which results from the fact that one can never completely
turn off the exchange coupling between any two given spins, and thus interactions of the two spins involved in
the SWAP operation with spins not involved in it will have an effect on the overall operation.  We also note that,
in the cases in which there is next-nearest-neighbor coupling, the infidelity appears to saturate to a finite,
non-zero, value, while it appears to decline to arbitrarily small values without next-nearest-neighbor coupling,
approaching zero as $J_\text{SWAP}$ grows.  This is not surprising, as, in the nearest-neighbor-only case, a large
$J_\text{SWAP}$ results in the other interaction terms becoming negligible compared to that between the two spins
being swapped, and the shorter time over which the pulse is applied means that there is less time for crosstalk
to have a significant effect on the fidelity.  On the other hand, when next-nearest-neighbor coupling is present, we
will never have all of the other interaction terms become negligible, and thus there will always be some noticeable
crosstalk effects, hence the saturation of the infidelity to a finite, nonzero, value.  We see similar effects
in the case in which the leftmost two spins start in the singlet state, as we show in Figs.\ \ref{fig:Nosj_Singlet_Plot_4}
and \ref{fig:Nosj_Singlet_Plot_6}.

\subsection{Noisy case}
\begin{figure}[htb]
	\centering
		\includegraphics[width=\columnwidth]{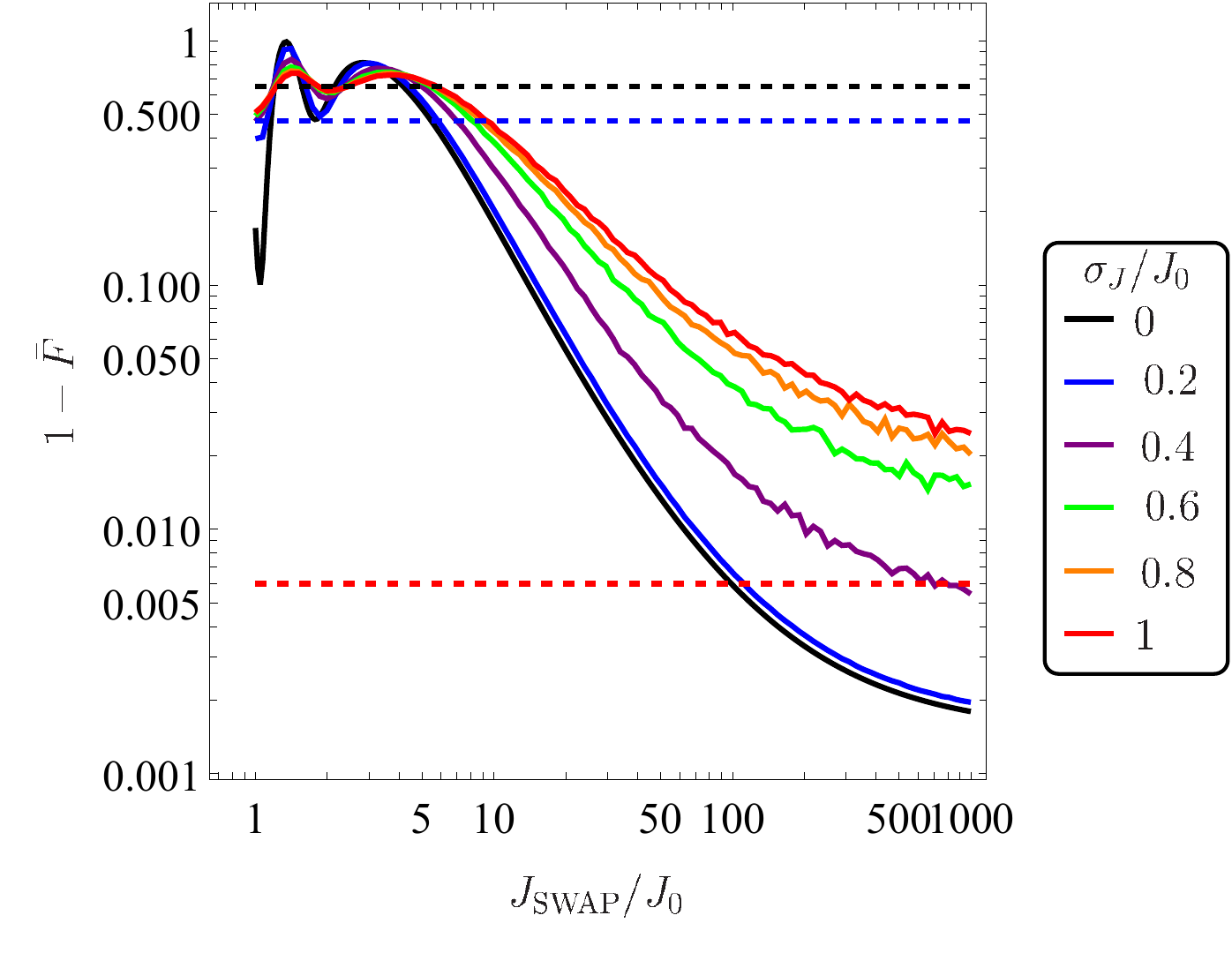}
	\caption{Plot of noise-averaged infidelity $1-\bar{F}$ of ``transporting'' a down spin from left to right and back
	again as a function of $J_\text{SWAP}/J_0$ for $4$ spins.  The system is initialized in the state, $\ket{\psi_{0,1}}=\ket{\downarrow\uparrow\uparrow\uparrow}$.
	We include a next-nearest-neighbor coupling $J'=0.01J$, dipole-dipole coupling, and with noise in $J$ with strength
	(i.e., standard deviation) $\sigma_J$.  The red dashed line corresponds to $99.9\%$ single SWAP fidelity, the blue dashed line to
	the experiment of Ref.\ \onlinecite{KandelNature2019}, and the black dashed line to the experiment of Ref.\ \onlinecite{SigillitoNPJQI2019}.}
	\label{fig:HighsjPlot_4}
\end{figure}

\begin{figure}[htb]
	\centering
		\includegraphics[width=\columnwidth]{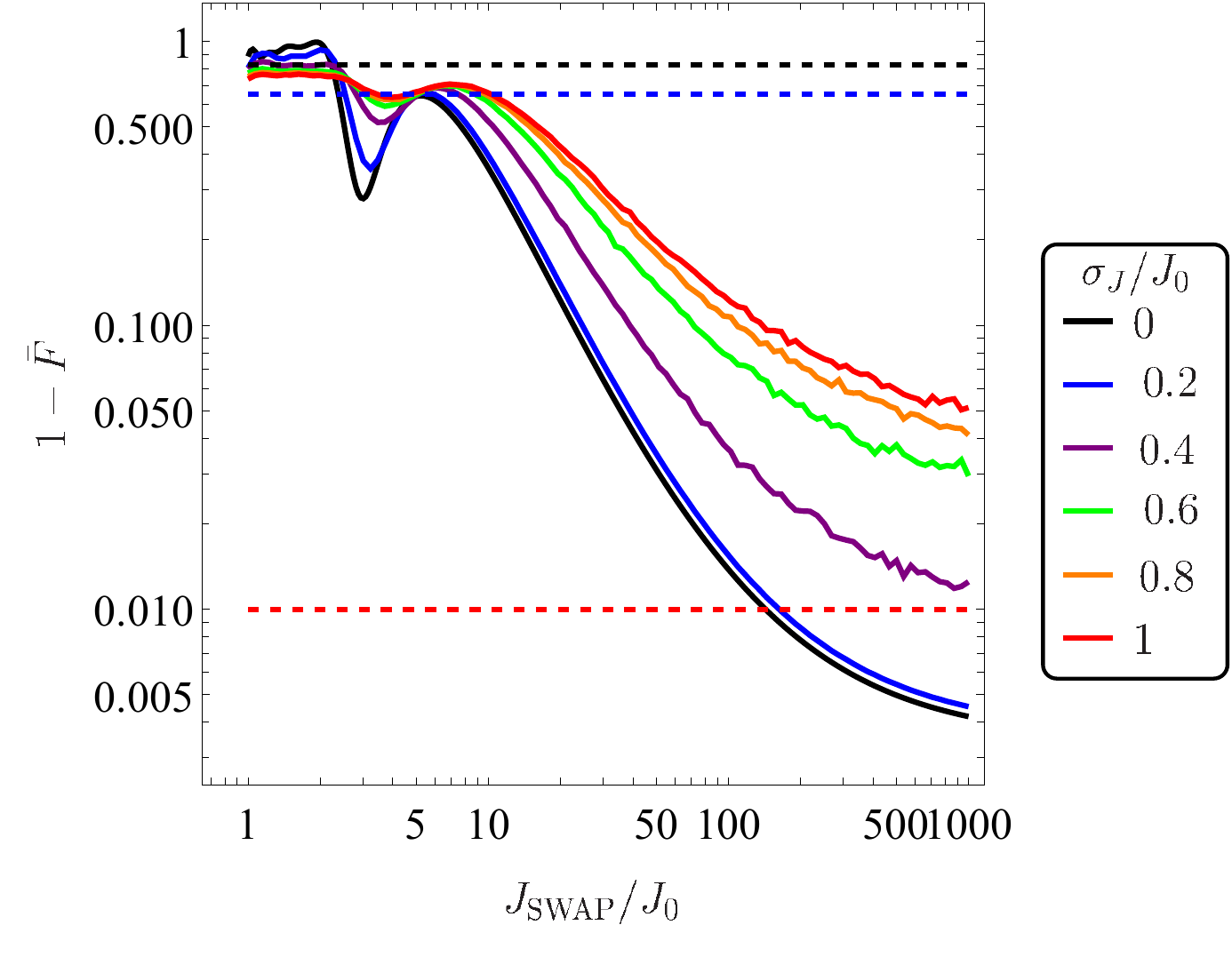}
	\caption{Plot of noise-averaged infidelity $1-\bar{F}$ of ``transporting'' a down spin from left to right and back
	again as a function of $J_\text{SWAP}/J_0$ for $6$ spins.  The system is initialized in the state, $\ket{\psi_{0,1}}=\ket{\downarrow\uparrow\uparrow\uparrow\uparrow\uparrow}$.
	We include a next-nearest-neighbor coupling $J'=0.01J$, dipole-dipole coupling, and with noise in $J$ with strength
	(i.e., standard deviation) $\sigma_J$.  The red dashed line corresponds to single SWAP $99.9\%$ fidelity, the blue dashed line to
	the experiment of Ref.\ \onlinecite{KandelNature2019}, and the black dashed line to the experiment of Ref.\ \onlinecite{SigillitoNPJQI2019}.}
	\label{fig:HighsjPlot_6}
\end{figure}

\begin{figure}[htb]
	\centering
		\includegraphics[width=\columnwidth]{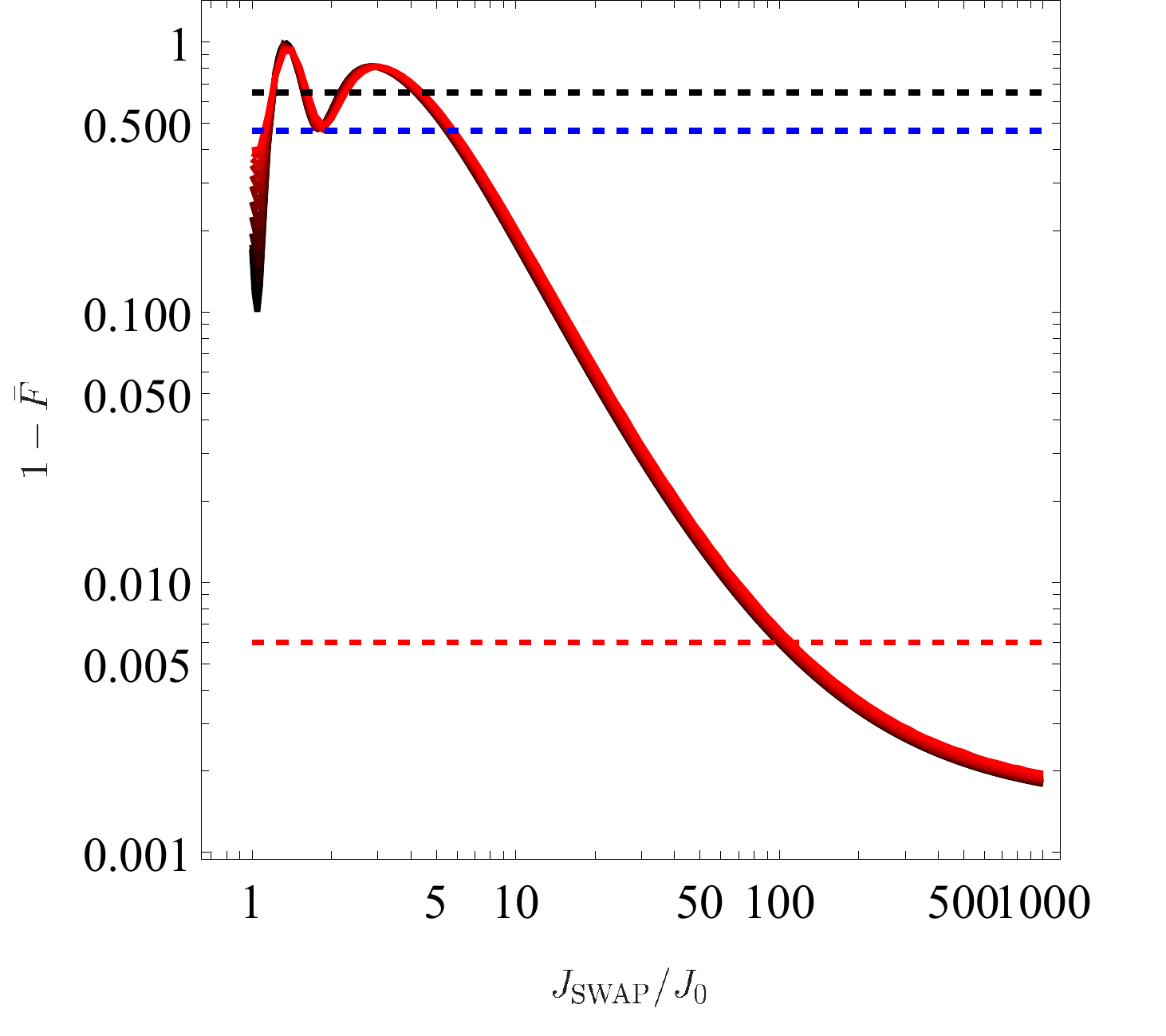}
	\caption{Plot of noise-averaged infidelity $1-\bar{F}$ of ``transporting'' a down spin from left to right and back
	again as a function of $J_\text{SWAP}/J_0$ for $4$ spins for low noise strengths, $0\leq\sigma_J\leq 0.2J$, in steps
	of $0.02J$.  The solid black curve corresponds to no noise, while the solid red curve corresponds to $\sigma_J=0.2J$.
	The system is initialized in the state, $\ket{\psi_{0,1}}=\ket{\downarrow\uparrow\uparrow\uparrow}$.
	We include a next-nearest-neighbor coupling $J'=0.01J$, dipole-dipole coupling, and with noise in $J$ with strength
	(i.e., standard deviation) $\sigma_J$.  The red dashed line corresponds to $99.9\%$ single SWAP fidelity, the blue dashed line to
	the experiment of Ref.\ \onlinecite{KandelNature2019}, and the black dashed line to the experiment of Ref.\ \onlinecite{SigillitoNPJQI2019}.}
	\label{fig:LowsjPlot_4}
\end{figure}

\begin{figure}[htb]
	\centering
		\includegraphics[width=\columnwidth]{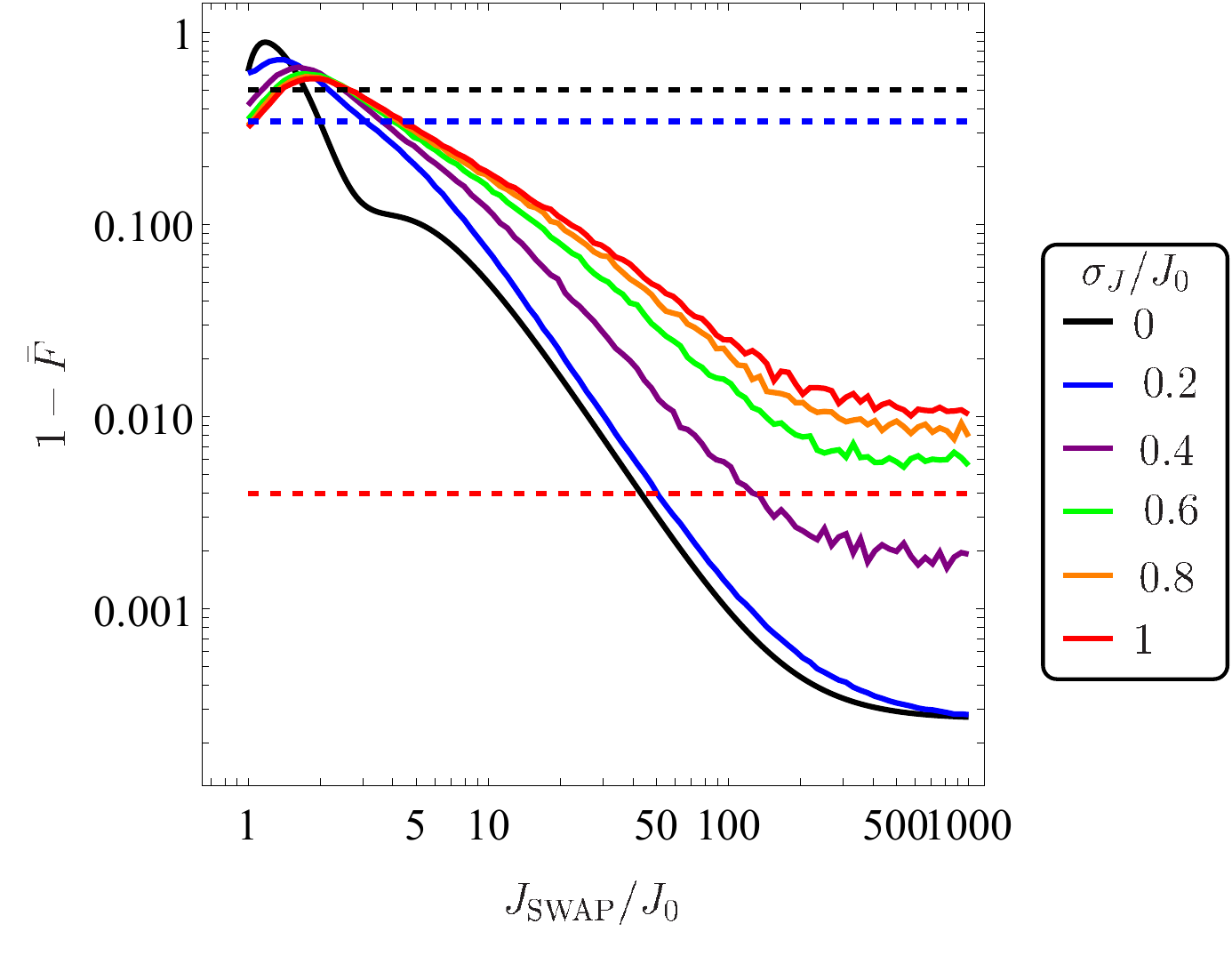}
	\caption{Plot of noise-averaged infidelity $1-\bar{F}$ of ``transporting'' one of two entangled spins from left to
	right and back again as a function of $J_\text{SWAP}/J_0$ for $4$ spins.  The system is initialized in the state,
	$\ket{\psi_{0,2}}=\ket{S}\otimes\ket{\uparrow\uparrow}$, where $\ket{S}=\frac{1}{\sqrt{2}}(\ket{\uparrow\downarrow}-\ket{\downarrow\uparrow})$.
	We include a next-nearest-neighbor coupling $J'=0.01J$, dipole-dipole coupling, and with noise in $J$ with strength
	(i.e., standard deviation) $\sigma_J$.  The red dashed line corresponds to $99.9\%$ single SWAP fidelity, the blue dashed line to
	the experiment of Ref.\ \onlinecite{KandelNature2019}, and the black dashed line to the experiment of Ref.\ \onlinecite{SigillitoNPJQI2019}.}
	\label{fig:Highsj_Singlet_Plot_4}
\end{figure}

\begin{figure}[htb]
	\centering
		\includegraphics[width=\columnwidth]{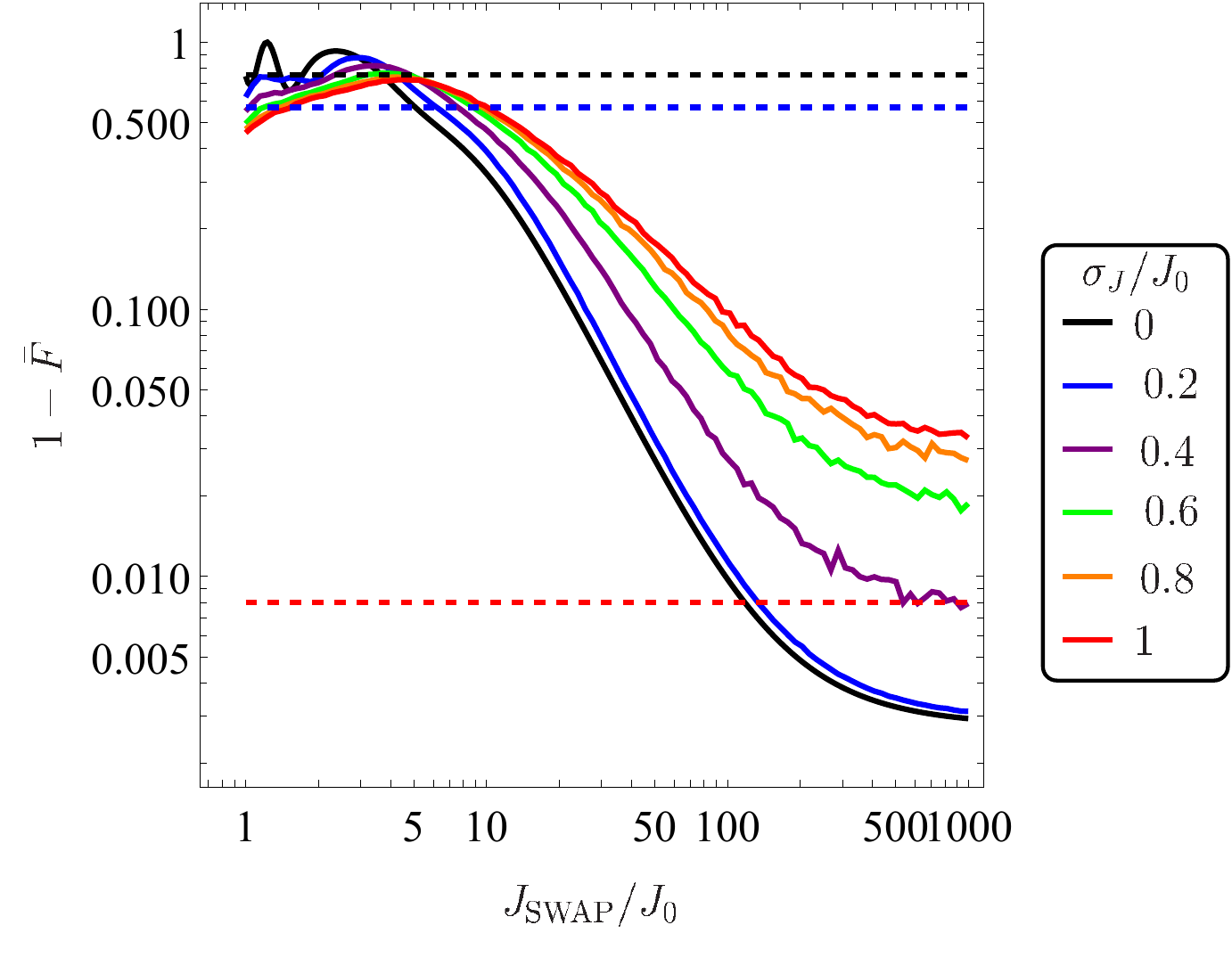}
	\caption{Plot of noise-averaged infidelity $1-\bar{F}$ of ``transporting'' one of two entangled spins from left to
	right and back again as a function of $J_\text{SWAP}/J_0$ for $6$ spins.  The system is initialized in the state,
	$\ket{\psi_{0,2}}=\ket{S}\otimes\ket{\uparrow\uparrow\uparrow\uparrow}$, where $\ket{S}=\frac{1}{\sqrt{2}}(\ket{\uparrow\downarrow}-\ket{\downarrow\uparrow})$.
	We include a next-nearest-neighbor coupling $J'=0.01J$, dipole-dipole	coupling, and with noise in $J$ with strength
	(i.e., standard deviation) $\sigma_J$.  The red dashed line corresponds to $99.9\%$ single SWAP fidelity, the blue dashed line to
	the experiment of Ref.\ \onlinecite{KandelNature2019}, and the black dashed line to the experiment of Ref.\ \onlinecite{SigillitoNPJQI2019}.}
	\label{fig:Highsj_Singlet_Plot_6}
\end{figure}

We now include quasistatic noise in our system.  We model this noise as follows.  The ``base'' values of the
nearest-neighbor exchange couplings $J_j$ are chosen from a Gaussian distribution $\sim e^{-(J-J_0)^2/2\sigma_J^2}$,
truncated so that all $J/J_0\in[0,\infty)$.  We assume here that $\sigma_J\propto J$, an assumption used in other
work on correction of noise-induced errors\cite{WangNatComms2012,ThrockmortonPRB2017}.  In practice, we implement the
linearity of $\sigma_J$ in $J$ by assuming that the actual coupling scales proportionately to the intended (i.e.,
without noise) coupling.  For each value of $\sigma_J/J_0$, we use $10,000$ realizations of noise and determine
the noise-averaged fidelity.  We plot the average infidelity for chains of $4$ and $6$ spins, with the system initialized
in the state $\ket{\psi_{0,1}}$, Figs.\ \ref{fig:HighsjPlot_4} and \ref{fig:HighsjPlot_6}.  We also show dashed lines
corresponding to the fidelities reported in the experiments of Refs.\ \onlinecite{KandelNature2019} and \onlinecite{SigillitoNPJQI2019},
as well as to $99.9\%$ fidelity.  We find that, for $\sigma_J\leq 0.2J$, the effect of noise on the fidelity is
very hard to discern visually; we illustrate this in Fig.\ \ref{fig:LowsjPlot_4}.
We see that, at least for large values of $J_\text{SWAP}$, the noise-averaged infidelity increases with $\sigma_J$, as
expected.  However, the relationship becomes more complicated for smaller values of $J_\text{SWAP}$---for some smaller
values of $J_\text{SWAP}$, the fidelity may actually slightly {\it increase} for increasing noise strength.  We have also
performed a similar calculation for the case in which we start with the state $\ket{\psi_{0,2}}$, and
we present the results in Figs.\ \ref{fig:Highsj_Singlet_Plot_4} and \ref{fig:Highsj_Singlet_Plot_6}.  We find similar
results as in the previous case.
Overall, our results imply that current experiments are unlikely to achieve the required $99.9\%$ fidelity needed in order
to implement error correction assuming the value of the next-nearest-neighbor exchange coupling that we assumed.  We also
find that the current experimental fidelities reported, both for four spin systems, correspond to higher noise strengths
than considered here, if we assume that $J_\text{SWAP}$ is tuned to the highest value considered, $J_\text{SWAP}=1,000J$.

We note that, in all of these calculations, we see nonmonotonic behavior in the fidelity as a function of $J_\text{SWAP}$
for small values $J_\text{SWAP}<5J_0$.  This is due to the fact that, for such small values of $J_\text{SWAP}$, the operation
takes a long time to execute, long enough that the additional oscillations induced by the crosstalk terms may go through (at
least approximately) a full period, thus resulting in partial cancellation of the effect.  For the values of $J$ for which
these oscillations appear, the nearest-neighbor crosstalk terms are the dominant contribution to these oscillations, as the
next-nearest-neighbor terms, even when included, will be much smaller than these, no more than about $0.05J_0$.  As a result,
the additional oscillations due to them will be much slower than those from the nearest-neighbor terms.

\section{Conclusions}
We investigated the fidelity of a sequence of SWAP operations performed on a chain of spins.  We considered three cases---one
in which only nearest-neighbor exchange coupling existed among the spins, one in which we included next-nearest-neighbor
coupling as well, and finally one in which we also added the magnetic dipole-dipole coupling (though it is very small) and
quasistatic noise in the exchange couplings.  In order to implement the SWAP gate between two neighboring spins, we increased
the coupling between those spins to a larger value $J_\text{SWAP}$ and maintained this value for a time $t=\pi\hbar/4J$; in
the cases in which we include next-nearest-neighbor coupling, we assume that these additional couplings involving the two spins
undergoing the SWAP operation increase in proportion to the nearest-neighbor coupling.  We also considered two initial states
for the system---a state in which the leftmost spin is down, while the rest are up (i.e., $\ket{\psi_{0,1}}=\ket{\downarrow\uparrow\uparrow\cdots\uparrow}$),
and one in which the leftmost two spins are in a singlet state while the rest are up (i.e., $\ket{\psi_{0,2}}=\ket{S}\otimes\ket{\uparrow\cdots\uparrow}$),
where $\ket{S}=\frac{1}{\sqrt{2}}(\ket{\uparrow\downarrow}-\ket{\downarrow\uparrow})$.  For each, we determined the fidelity
of a sequence of SWAP gates that moved the state of the leftmost spin (starting in the state $\ket{\psi_{0,1}}$) or the
second spin from the left (starting in the state $\ket{\psi_{0,2}}$) all the way to the right and then back again as a
function of $J_\text{SWAP}$ for chains of $4$ and $6$ spins.

We found that, even in the case in which there is just nearest-neighbor coupling, the fidelity of this sequence of operations
is reduced due to crosstalk, though it tends to unity as we increase $J_\text{SWAP}$.  If we introduce next-nearest-neighbor
coupling as well, then we find that the fidelity saturates at a value less than unity, no matter how large we make $J_\text{SWAP}$.
This is due to the fact that we assume that the next-nearest-neighbor couplings involving the spins undergoing the SWAP operation
increase in proportion to the nearest-neighbor coupling, so that there is always significant crosstalk.  We then added in the
magnetic dipole-dipole interaction and quasistatic noise.  We found that, for large values of $J_\text{SWAP}$, the fidelity
decreased monotonically for increasing noise strength, as expected, but that it could slightly {\it increase} with increasing
noise strength for some smaller values of $J_\text{SWAP}$.  In each case, we show what level of noise we could expect in different
experiments given the reported fidelities in each.

Our results imply that, while current experiments are unlikely to achieve the necessary $99.9\%$ fidelity required for error correction
techniques to be employed, it is still possible to achieve such a fidelity if noise in the exchange couplings were to be reduced.
Despite this, methods for combating the effects of noise and crosstalk are still of great interest.  One means by which higher fidelities
may be achieved is through error-correcting pulse sequences that cancel out the effects of error-inducing terms to a given order,
similar to those described in Refs.\ \onlinecite{WangNatComms2012} and \onlinecite{ThrockmortonPRB2017}.  Another method is to reduce
the noise in the exchange couplings, which could be achieved by reducing noise in the voltage on the gates used to define the quantum
dots in experimental systems.  It should be noted, however, that our conclusions assume that the next-nearest-neighbor couplings are
$1\%$ of the nearest-neighbor couplings; if this percentage is lower, then higher fidelities will be possible.  We should also note
that we deal here solely with errors due to noise in the exchange couplings, presumably itself due to noise in the voltage sources
used to define the quantum dots that the electrons reside in; another potential source of error, the treatment of which is beyond the
scope of this work, are sources intrinsic to the semiconductor system, such as spin-orbit coupling\cite{LiPRA2017}.

\acknowledgments
This work was supported by the Laboratory for Physical Sciences.

\end{document}